\documentstyle[aas2pp4]{article}

\def\'#1{{\ifx#1i{\accent"13\i}\else{\accent"13#1}\fi}}

\def\dphidt{d \Phi /dt}
\def\etal{{\rm et. al}}

\def\ekin{{\cal E}_{\rm kin}}
\def\eint{{\cal E}_{\rm th}}
\def\Ei{E_{\rm i}}
\def\Eg{E_{\rm g}}

\def\gamcr{\gamma_{\rm cr}}
\def\gameff{\gamma_{\rm eff}}
\def\gamg{\gamma_{\rm g}}
\def\gamt{\gamma_{\rm t}}

\def\kms{km s$^{-1}$}

\def\rhoc{{\rho_{\rm c}}}

\def\rhot{\rho_{\rm T}}

\def\PTERM{{P_{\rm th}}}

\def\va{v_{\rm a}}

\def\VS{V\'azquez-Semadeni}

\def\x{{\bf x}}
\def\XP{1/2\oint_S{x_i P_{\rm th}\hat n_i dS}}
\def\xp{{\cal T_{\rm th}}}
\def\XRUU{{1/2\oint_S{x_i \rho u_i u_j \hat n_j} dS}}
\def\xruu{{\cal T_{\rm kin}}}

\def\alamenos#1{$^{-#1}$}
\def\ala#1{$^{#1}$}

\begin{document}

\title{Clouds as turbulent density fluctuations. Implications for
pressure confinement and spectral line data interpretation}

\author{Javier Ballesteros-Paredes$^1$, Enrique
V\'azquez-Semadeni$^1$, and John Scalo$^2$}

\affil{$^1$Instituto de Astronom\'\i a, Universidad Nacional 
Aut\'onoma de M\'exico \\ Apdo. Postal 70-264, 04510 M\'exico D.F.,
M\'{e}xico.  {e-mail: {\tt javier,enro@astroscu.unam.mx}} }

\affil{$^2$Astronomy Department, University of Texas, Austin, TX
78712-1083.\\ {e-mail: {\tt parrot@astro.as.utexas.edu}} }

\begin{abstract}
We examine the idea that diffuse and giant molecular clouds and their
substructure form as density fluctuations induced by large scale
interstellar turbulence. We do this by closely
investigating the topology of the velocity, density and magnetic fields
within and at the boundaries of the clouds emerging in high-resolution
two-dimensional simulations of the ISM including self-gravity,
magnetic fields, parameterized heating and cooling and a simple model
for star formation. We find that the velocity field is
continuous across cloud boundaries for a hierarchy of clouds of
progressively smaller sizes. Cloud boundaries defined by a
density-threshold criterion are found to be quite arbitrary, with no
correspondence to any actual physical boundary, such as a density
discontinuity. Abrupt velocity jumps are coincident with the density
maxima, indicating that the clouds are formed by colliding gas streams.
This conclusion is also supported by the fact that the volume
and surface kinetic terms in the Eulerian Virial Theorem for a cloud
ensemble are comparable in general.
The topology of the magnetic field is also suggestive
of the same process, exhibiting bends and reversals where the gas
streams collide. However, 
no unique trend of alignment between density and
magnetic features is observed. Both sub- and super-Alfv\'enic motions
are observed within the clouds in the simulations.

In the light of these results, we argue that thermal pressure equilibrium is
irrelevant for cloud confinement in a turbulent medium, since inertial
motions can still distort or disrupt a cloud, unless it is strongly
gravitationally bound. Turbulent pressure confinement appears
self-defeating, because turbulence contains large-scale motions which
necessarily distort Lagrangian cloud boundaries, or equivalently cause
flux through Eulerian boundaries.

We then discuss the compatibility of the present scenario with
observational data. We find that density-weighted velocity histograms
are consistent with observational line profiles of comparable spatial
and velocity resolution, exhibiting similar FWHMs and similar
multi-component structure. An analysis of the regions contributing to
each velocity 
interval indicates that the histogram ``features'' do not come from
isolated ``clumps'', but rather from extended regions throughout a cloud,
which often have very different total velocity vectors. 

Finally, we argue that the scenario presented here may be also
applicable to small scales with larger densities (molecular clouds and their
substructure, up to at least $n \sim 10^3$--$10^5$ cm${-3}$), and
conjecture that quasi-hydrostatic 
configurations cannot be produced from turbulent fluctuations unless
the thermodynamic behavior of the flow becomes nearly adiabatic. We
demonstrate, using appropriate cooling rates, that this will not occur
except for very small compressions ($\lesssim 10^{-2}$ pc) or until
protostellar densities are reached for collapse.

\end{abstract}

\section{Introduction}\label{intro}

Theoretical models of interstellar clouds and clumps most frequently
assume static or stationary configurations, the clouds being either
confined by the external pressure (\cite{maloney88}; Bertoldi \&\
McKee 1992, herafter BM92), or
in ``virial equilibrium'' between 
its self-gravity and some form of internal energy, be it thermal (e.g.,
\cite{chieze87}), micro-turbulent (e.g., \cite{chandra51};
\cite{bona_etal87}; \cite{leorat_etal90}; \cite{VS_gaz95}) or magnetic
(e.g., \cite{shu_etal87}; \cite{mousch87}, \cite{MG88b}), forming out
of instabilities or coagulation of smaller clouds (see \cite{elm93a}
for a review). Nevertheless, the interstellar medium (ISM) is well
known to be highly violent (e.g., \cite{McCray_snow79}), and recently it has
become increasingly accepted that it is turbulent throughout (e.g.,
\cite{scalo87}; see also the volume ``Interstellar Turbulence''
[\cite{franco_carram98}]). In such a medium, clouds may naturally form as
turbulent density fluctuations as well.

Within this dynamic, turbulent framework, it is important to reconsider
the meaning and feasibility of pressure cloud confinement, either
thermal or turbulent. This is most adequately done in the context of the
Virial Theorem (VT), which is a very useful tool for describing the
balance between all the physical agents in molecular clouds.
Customarily, the VT is considered in a Lagrangian frame, i.e., a frame
moving with the flow. However, there are circumstances when
one would prefer to use an Eulerian description. For example, a
Lagrangian description is clearly not well suited for describing the
propagation of a wave, in which the fluid does not move along with the
perturbation, while in the case of an isolated object a Lagrangian
description fits most naturally (see also \VS, Passot \& Pouquet 1996,
hereafter \cite{VSPP96}). In the interstellar case, if significant mass
exchange occurs between a cloud and its surroundings, such as an
accreting cloud, an Eulerian (fixed in space) may be preferable since a
Lagrangian boundary would undergo severe distortions. Parker (1969)
derived an Eulerian form of the Virial Theorem (EVT) in tensor form,
but he explicitly neglected the mass flux through the surface of the
cloud. More recently, McKee \& Zweibel (1992) (hereafter \cite{MZ92})
have written all the terms entering the EVT:

\begin{eqnarray}
{1\over 2} \ddot{I}_E = 2 \biggl(\eint + \ekin - 
\xp -  \xruu  \biggr) + {\cal M} + \nonumber \\
{\cal W} - {1\over 2} {d\Phi\over dt},
\label{EVT}
\end{eqnarray}
where $I_E\equiv \int_V \rho r^2 dV$ is the moment of inertia of the
cloud, $\eint\equiv 3/2\int_V {P_{\rm th} }dV$ is the thermal energy,
with $P_{\rm th}$ being the thermal pressure; 
$\ekin\equiv 1/2\int_V \rho u^2 dV$ is the kinetic energy,
$\xp\equiv\XP$ is the surface thermal term, $\xruu\equiv\XRUU$ is the
surface kinetic term, $ M\equiv 1/8\pi\int_V B^2 dV + \int_S
x_i T_{ij} \hat n_j dS$ is the magnetic term, with $T_{ij}$ the
Maxwell stress tensor, $ W \equiv \int_V \rho x_i g_i dV$ is the
gravitational term and $ \Phi \equiv\oint_S \rho x^2 u_i \hat n_i
dS$ is the flux of moment of inertia through the surface of the
cloud. Sums over repeated indices are assumed unless otherwise stated.

Frequently, the surface terms are neglected altogether, especially in
observational work (\cite{larson81}; \cite{Torrelles83}; \cite{MG88a};
\cite{ful_mye92}), although it is also a common
practice in theoretical studies, as a consequence of assuming isolated
clouds (e.g., \cite{Chandra53}; \cite{parker69},
\cite{parker79}). The most notable
exception is the thermal pressure surface term $\xp$, which is
frequently invoked for ``pressure confinement'' (e.g. \cite{McCrea57};
\cite{KetoMyers86}; \cite{maloney88};
\cite{BM92}; \cite{mcl_pud96}; \cite{yonekura_etal97}). 
By analogy,  \cite{MZ92} have
considered the possibility of  
turbulent pressure confinement by means of the term $\xruu$. However, in
the present paper we will argue that both kinds of pressure confinement
require idealized conditions that are not likely to be realized in
the actual ISM. If the interstellar medium (ISM) is globally
turbulent, the velocity 
field is in general locally nonzero, and thermal
pressure balance may be irrelevant, since a cloud can still
undergo deformation due to the inertial motions, which will induce a
nonzero RHS via the kinetic terms $\ekin$, $\xruu$, and $d
\Phi/dt$. 

Concerning turbulent pressure confinement, its feasibility requires the
underlying assumption that the turbulence be microscopic, so that it
can be considered isotropic compared to the scale of the cloud.
However, turbulence is inherently a multi-scale phenomenon, and is
expected to contain excitation at scales comparable or even larger to
that of the cloud. It is in this context that the clouds may be
considered as density fluctuations produced by larger-scale compressive
turbulent motions (e.g., \cite{hunter79}; \cite{larson81};
\cite{hunt_fleck82}; \cite{hunter_etal86}; \cite{elm93b}; \VS, Passot
\&\ Pouquet 1995, hereafter \cite{VSPP95}).  Such clouds are then
turbulent fluctuations, and can either rebound, fragment or collapse
depending on the compressive energy available, the cooling ability of
the flow, the topology of the velocity and magnetic fields, the
production of internal turbulent motions and a number of instability
mechanisms (e.g., \cite{E&E78}; Vishniac 1983, 1994;
\cite{hunter_etal86}; Tohline, Bodenheimer \&\ Christodoulou 1987;
\cite{Stevens_etal92}; \cite{elm93b}; \cite{VSPP96};
\cite{Korn_Sca98}).

Note that in this scenario, clouds and clumps need not be
in a static or quasi-static equilibrium at any time. Moreover, the
fact that clouds and clumps are observed to be internally turbulent
suggests that the turbulence has not been completely dissipated
inside.  In fact, it is known that shocks
running obliquely through a density gradient will produce further
internal turbulence (\cite{klein_etal95}; \cite{Korn_Sca98}).  

In this paper, we analyze the density, velocity and magnetic fields in
two-dimensional numerical simulations of turbulence in the ISM in order
to estimate the importance of the transport processes through the cloud
boundaries. We start in \S \ref{kin_sec} by discussing the role of
the  kinetic terms in the Virial Theorem and the implications for
pressure confinement. Then, after briefly describing the numerical
simulations (\S \ref{simulations}), we proceed to a description of our
results (\S\ \ref{results}), starting with a discussion of the
density, velocity and magnetic features that arise, and their spatial
correlations (\S\ref{topology}). Then we present an evaluation of 
the surface and volume kinetic terms in the virial
theorem, suggesting that their similarity may be indicative that both
types of terms are representative of the same phenomenon (\S\
\ref{eval_surf}). The super-
or sub-Alfv\'enic character of the motions is discussed in
\S\ref{super_sub}. In \S\ \ref{comp_obs} we present several
comparisons with known observational data, such as 
line profiles
observed in cloud complexes (\S\ref{spectra}), the topology of the
magnetic field (\S\ \ref{B_comparison}), and the lifetimes of the
clouds (\S\ \ref{lifetimes}).
In \S\ \ref{discussion} we then discuss several issues and possible
caveats, like whether the scenario proposed here can be applicable to
smaller, denser scales, the low probability of forming quasistatic clumps in a
turbulent medium, and the effect of the dimensionality of our
simulations. Finally, we give a summary and some conclusions in
\S\ref{conclusions}.

It is interesting to note that the present
paper may be regarded as theoretical justification for the suggestion made
over 45 years ago by \cite{chandra_munch52} that an alternative to
the picture that visualizes the interstellar medium as consisting of a
distribution of discrete clouds might be necessary.

\section{Kinetic Terms in the Eulerian Virial Theorem}\label{kin_sec}

As is well known (e.g., \cite{parker69}; \cite{MZ92}), the EVT can be
obtained from the conservative form of the momentum equation,
\begin{equation}
{\partial (\rho u_i)\over\partial t} + {\partial (\rho u_i u_j)\over
  \partial x_j} = \sum_k F_k
\label{mom}
\end{equation}
(where $F_k$ generically represents the various force densities
involved in the problem) by dotting it with the position vector \x\
and integrating over a volume $V$ fixed in space. 
We remind the reader that the full Virial Theorem, eq.\
(\ref{EVT}), holds always because it is a direct consequence of the
momentum equation, although this does not necessarily imply that the
system is in equilibrium or, in particular, that the linewidth of an
interstellar cloud is directly related to its mass and radius
(\cite{maloney88}). A detailed analysis of the VT in our 
simulations will be presented elsewhere (\cite{BPetal98}). 

In the present paper we focus on the relative importance of the surface and
volumetric kinetic terms, and their role in shaping the clouds. It can
be easily checked that, in the EVT (eq.\ [\ref{EVT}]), the kinetic
energy ${\cal E}_{\rm 
kin}$, the kinetic surface term ${\cal T}_{\rm kin}$, and the
time-derivative terms $\ddot I_E$ and $\dphidt$, originate from the LHS
in the momentum equation (eq.\ [\ref{mom}]).
While the meaning of the kinetic energy term $\ekin$ is obvious, 
the kinetic surface term $\xruu$ has two possible interpretations.
One is as half the flux of the instantaneous rate of change of the moment
of inertia $\rho x_i u_i$ through the cloud's surface. Alternatively,
it can be interpreted as the sum of the ram 
pressure plus the kinetic stresses, both evaluated at the surface of
the cloud:
\begin{eqnarray}
\oint_S x_i \rho u_i u_j \hat{n}_j dS = \oint_S x_i \rho u^2 \hat{n}_i
dS + \nonumber \\ \oint_S x_i \rho u_i u_j \hat{n}_j dS,
\label{surface_sum}
\end{eqnarray}
where $i\neq j$ in the last surface integral. The first term on the RHS
of the eq.\ (\ref{surface_sum}) is analogous to the thermal pressure
surface term. The second term, on the other hand, reflects the fact
that the turbulent motions are not isotropic, giving off-diagonal
contributions to the total pressure tensor $\Pi_{ij}= \PTERM
\delta_{ij} + \rho u_i u_j$ (\cite{land_lift}). This exhibits in a
clear way the difference between the 
isotropic nature of the thermal pressure and the anisotropic nature of
the ``turbulent'' pressure.

Another important difference between the thermal and ``turbulent''
pressures lies in the fact that the latter is in general expected to
involve motions of scales  comparable to that of the cloud, as
mentioned in \S\ref{intro}. But since the velocity field carries
mass with it, such large-scale motions necessarily imply a mass
flux across fixed, Eulerian cloud boundaries. Conversely, in a
Lagrangian description, such turbulent motions should lead to severe
distortions of the Lagrangian cloud boundary, possibly giving it a
fractal character (e.g., Scalo 1990; \cite{falg_etal91}).

The exception to the scenario depicted above is the case of strongly
self-gravitating clumps in (magneto-) hydrostatic equilibrium, from
which the gas cannot escape once it has been captured (``decouples''
from the intercloud medium). In this case, the cloud should be bounded
by an accretion shock and, although strictly speaking there is flux
across the shock, the accreted material is not mixed far beyond the
shock into the cloud's ``body''. However, it seems to us that the
question as to whether such quasi-hydrostatic configurations can actually
be produced in a turbulent ISM remains open, as discussed in \S\ \ref{gamma}.

\section{Numerical Simulations}\label{simulations}

In the following sections, we discuss the topology of the density,
velocity and magnetic fields in 2D numerical simulations of the ISM
based on the model of Passot, \VS\ \&\ Pouquet (1995) (hereafter
\cite{PVSP95}), which represents 1~kpc$^2$ of the ISM on the Galactic
plane, centered at the solar Galactocentric distance. We refer the
reader to that paper for the equations and parameters of the model,
which includes self-gravity, magnetic fields, parameterized cooling and
diffuse heating, the Coriolis force, large-scale shear, and
parameterized localized stellar energy input due to ionization
heating.  The parameterized cooling is as in \cite{chi_breg88}, who
fitted piecewise power laws to the standard cooling calculations of
\cite{dalg_Mc72} and \cite{ray_etal76}. As discussed in \cite{VSPP96},
the net result of the combined cooling and diffuse heating laws is that
the flow effectively behaves as a polytrope, with an effective
polytropic exponent $\gameff$ which is piecewise constant over the
temperature range spanned by the flow.  Alternatively, it was
shown in \cite{VSPP95} that, for the heating and cooling functions
used there, the thermal equilibrium values of the
temperature and pressure can be expressed as functions of the density
in the range 100 K $< T <10^5$ K for non star-forming regions. Therefore,
$\gameff$ can also be expressed as a piecewise constant over the
corresponding density range. For the fiducial values adopted in the
model (\cite{PVSP95}), $\gameff$  takes the values (\cite{VSPP96})
\begin{equation}\label{table_gamma}
\gameff=\cases{
	0.25 &  $1.57 < \rho\ (100< T < 2000)$ \cr
	0.   &  $0.39 < \rho \leq 1.57\ (2000 \leq T < 8000)$\cr
	0.48 &  $3.15 \times 10^{-3} < \rho \leq 0.39\ (8000 \leq T < 10^5)$\cr
	3.3  &  $4.28 \times 10^{-2} < \rho \leq 3.15 \times 10^{-3}$\cr
             &   $(10^5 \leq T < 4 \times 10^7)$,\cr 
}
\end{equation}
where densities are in units of cm$^{-3}$ and temperatures in Kelvins.
In eq.\ (\ref{table_gamma}), equilibrium temperature ranges equivalent to
the specified density ranges are indicated in parentheses. In the
simulations, stellar ionization heating is modeled by means
of a local heating source which is turned on if the density exceeds a
critical value $\rhoc=30\langle \rho \rangle$, where the brackets
denote an average over the whole volume of the simulation. Note that
temperatures over a few $\times 10^4$ K are never reached in the
simulations, because supernovae are not included, since these require
non-trivial modifications to the code (\cite{gazol_pass98}). The maximum 
densities reached in the simulations are $\sim 100$ cm$^{-3}$, with $T
\sim 100$ K.

The simulations we use are refinements of the \cite{PVSP95} model at a
higher resolution (800$\times$800 grid points), presented in
V\'azquez-Semadeni, Ballesteros-Paredes, \& Rodr\'iguez  (1997), hereafter
\cite{VSBPR97}.  The stellar energy injection maintains the turbulence
in the medium, promoting further cloud formation. Nevertheless, we turn
off star formation shortly before the time at which the data are
analyzed, in order to allow for the largest possible density gradients,
since otherwise the stellar heating prevents the density from reaching
values significantly larger than $\rhoc$, by causing the neighboring
gas to re-expand.  The various physical quantities are in units of
$\rho_0=1$ cm$^{-3}$, $u_0=11.7$ km s$^{-1}$, $T_0=10^4$ K and
$B_0=5\mu$G (see \cite{PVSP95}).

We have developed a cloud-finding numerical algorithm in physical space
(as opposed to the position-velocity space of observational data) for
the density data from the simulations (\cite{BPVS95}). A cloud is
defined as a connected set of points (pixels) whose densities  are
larger than a given threshold $\rhot$. The clouds obtained through this
procedure are clearly quite arbitrary, since as $\rhot$ is increased
the boundary of the cloud simply moves ``inwards''. However, eventually
an increase in $\rhot$ will lead to a more qualitative change in which
a given cloud splits into two or more ``children'', in a similar manner
to the ``structure trees'' used by \cite{hou_sca92}. This procedure is
in a sense analogous to performing observations using different tracers
sensitive to different density ranges. Alternative cloud
identification algorithms based on laocating density maxima (e.g.,
\cite{will_etal94}) were not investigated, but we
do not expect that such procedures would affect the qualitative nature
of our results, which, as will be seen, appear to follow quite
generally from the physics of the problem.

Given the above algorithm for defining the clouds, the kinetic terms
entering the VT are then evaluated numerically for each cloud, having
previously substracted its bulk mass-averaged velocity, defined as
\begin{equation}
\langle u_i \rangle_{\rm mass} \equiv {\int_V \rho u_i dV  \over \int_V
\rho dV},
\label{umass}
\end{equation}
in order to measure only the contribution associated
to the fluctuations. We refer to this as measuring the velocities in
the ``cloud frame''.

\section{Results and Discussion}\label{results}

\subsection{Density and velocity field topology}\label{topology}

In figs.\ \ref{family}a, b and c we show three hierarchical levels of
the density field (gray scale), with the corresponding velocity fields
(arrows). In fig.\ \ref{family}c we further denote super- and
sub-Alfv\'enic velocities with black and white arrows, respectively
(see \S\ \ref{comp_obs}), where the Alfv\'en speed is defined as
$\va^2=B^2/\rho$ in the code's units.

Figure \ref{family}a shows a snapshot of the simulation at $t=7.2
\times 10^7$yr. The density field ranges from $\sim 0.01$ to $\sim
100$cm\alamenos 3.  Figs.\ \ref{family}b and c show two subsequent
hierarchical levels of clouds, obtained at $\rhot=4$ and 8,
respectively. The sizes 
of the boxes shown in the three panels are 800, 215 and 45 pixels per
side (1 pixel = 1.25 pc). We refer to the cloud in fig.\ \ref{family}b
as the ``parent'' cloud, and to that in fig.\ \ref{family}c as the
``child'' cloud. Note
that only fig.\ \ref{family}c shows the velocity field in the cloud frame. 

Inspecting the plotted density and velocity fields, we observe that
the fluid velocity at the cloud boundaries is in general continuous,
at any level of the hierarchy. Sharp changes in the velocity field
(shocks)\footnote{In the simulations, shocks are sharp gradients
extending over 3 to 5 pixels, due to the action of viscosity (\cite{VSBPR97})}
are in general oblique and tend to occur mostly at the centermost
parts of the clouds, where the highly filamentary density field
exhibits ``ridges''. This indicates that the density features are
produced by colliding streams, a fact that can be seen more accurately
in figs. \ref{cut}a, b and c. These show cuts,
along the $x$- (at $y=514$, fig. \ref{cut}a) and $y$- (at $x=602$ and
$x=614$, figs. b and c respectively) axes, of the density $\rho$ (solid
line), and the $x$- and $y$-components of the velocity, $u_x$ (dotted
line) and $u_y$ (long dashed line) for a region containing the clouds
of figs.\ \ref{family}b and c. The axes in these plots indicate pixel
number. For added clarity, we show in fig.\ 
\ref{div} the lines along which these cuts are taken. 

The central
parts (3-5 pixels) of the density peaks are seen to correspond in
general to large negative longitudinal gradients $\partial u_i/\partial
x_i$ (no sum over $i$) of the velocity, i.e., shocks.
However, the clouds (i.e., large density values) extend beyond the
shocks for many more pixels, where the velocity component exhibits
plateaus with superimposed small-scale structure. These plateaus are
indicative of the approaching streams (extended inward motions). As an
example, consider
the peaks labeled ``1'' and ``2'' in fig.\ \ref{cut}a. Peak 1
corresponds to a shock at $x=602$, and is flanked by an expansion wave
($\partial u_x/\partial x >0$) on the left, and by a plateau with $u_x
<0$ on the right, although the plateau itself contains
substructure. This peak also corresponds to a shock along the
$y$-direction (fig.\ \ref{cut}b).  Interestingly, peak ``2'' in
fig.\ \ref{cut}a does not seem to correspond to a shock along the
$x$-direction, but in fig.\ \ref{cut}c it can be seen that it is caused
by a shock in the $y$-direction. 
In order to further illustrate this point, in fig.\ \ref{div} we show
the same cloud as in fig.\ \ref{family}c, but showing the density field
as a gray-scale image and the divergence $\partial u_i/\partial x_i$ of
the velocity field with countours. Local density maxima clearly
coincide with local (negative) minima of the divergence.

It is worth noting that in figs.\ \ref{cut}a, b and c, larger-scale
density features are also seen to correspond to large-scale
negative gradients of the velocity field, although with much
substructure. See, for example, the large cloud complex (multiple
density peak) extending from $580 \lesssim x \lesssim 680$ in fig.\
\ref{cut}a. The very largest-scale inward flows are probably of
gravitational rather than turbulent origin (e.g., \cite{VSP98}),
because the simulations do not include supernovae which could
induce turbulent motions at those scales, but in the real ISM both
cases are likely to occur.

There also appears to be a certain amount of spatial
correlation between density and vorticity features. In figure
\ref{vort} we show the vorticity field (contours) superposed onto the
density field (gray-scale) for the cloud of fig.\ \ref{family}c. In this
case, the correlation between local 
maxima of the density field and maxima (or minima) of the vorticity
field is not as tight, the critical points of the vorticity being
shifted by a few pixels with respect to the density peaks.
Nevertheless, there is still a clear correspondence between vorticity
and density features. This correlation can be understood because, as
mentioned above, the stream collisions are in general oblique. It is possible
then that the detailed structure inside the clouds consists of shocks
and tangential discontinuities.
Moreover, the shocks are generally curved and are encountering density
variations in the pre-shock gas, so vorticity
production is expected behind them (\cite{hayes57}; \cite{PP87};
\cite{Korn_Sca98}).

The continuity of the velocity field through the cloud boundaries has
an interesting subjective implication. The definition of a cloud by a
density threshold is actually seen to be rather arbitrary, since the
clouds do not have sharp physical boundaries. Thus, as long as the
change in the density
threshold does not imply a ``splitting'' of a cloud into its offspring,
it only amounts to considering the same cloud out to different
distances from the maximum, but without there being any clear ``edge''
to the cloud.
Note that this need not be in conflict with observations suggesting
that molecular clouds have sharp edges (\cite{blitz91}),
since in this case the observed boundaries may just reflect a 
transition from molecular to atomic gas, without a discontinuity in the
gas mass density (e.g., \cite{goldsmith87}, \cite{blitz91}). 

\subsection{Virial surface terms and pressure confinement}
\label{eval_surf}

The velocity fields observed in the simulations indicate a strong flux
of the relevant dynamical quantities through the cloud boundaries at
all scales. To quantify this, in fig.\ \ref{kinetic_terms} we show a
plot of the kinetic energy $\ekin$ vs.\ the kinetic surface term
$\xruu$ for the parent complex shown in fig.\ \ref{family}a and all its
daughter clouds resulting from setting $\rhot=$4, 8, 11 and 16, and
with areas larger than 300 pixels. The latter requirement is imposed in
order to avoid including clouds so small that they are strongly
influenced by viscous and diffusive numerical effects. The clouds shown in
figs.\ \ref{family}b and c are represented with a different symbol in
fig.\ \ref{kinetic_terms} for identification. Note that both kinds of
virial terms are computed in the cloud frame.

It is seen that in general the surface kinetic virial term is of the
same order of magnitude as the volume term, i.e., the kinetic energy
contained in the cloud's volume. This means that both terms are of
similar importance in shaping and supporting the clouds. At a
more detailed level, this seems to be a reflection of the continuity of the
velocity field. The flow is entering the clouds, and shocking at their
innermost regions. Thus, the two terms refer to essentially the same
process, only measuring it at different places (one over the cloud's
volume, the other at the cloud's boundary but weighted by the distance
to the center of mass), explaining their similar values. This is
analogous to the well known property of the surface and volume thermal
pressure terms, that if the pressure is constant the two terms cancel.

It is important to remark that the similarity between these two terms
does not imply that the cloud is ``confined'' by turbulent pressure, in
the sense that this is not a static configuration. First, the two
terms do not cancel each other exactly, leaving a net contribution
available for shaping the cloud (the contribution $\ekin-\xruu$ to
$d^2I/dt^2$). Second, it is important to note that the points in fig.\
\ref{kinetic_terms} lie sometimes above and sometimes below the line
$\ekin=\xruu$, indicating that for some clouds the contribution to
$d^2I/dt^2$ is positive, but negative for others. This implies that the
kinetic terms sometimes provide net ``support'' and sometimes cause net
compression, although in general the clouds are expected to suffer
distortions that cannot be classified as either one of the above.
Thus, the clouds
are evolving with time, continuously changing their volume and shape,
since they are the density fluctuations produced by the 
turbulence\footnote{See the accompanying video to \cite{VSPP95} for a
non-magnetic example. An mpeg video of a fully MHD simulation from
\cite{PVSP95}, among others, can be seen at {\tt
http://www.astroscu.unam.mx/turbulence/movies.html}}. 

The thermal pressure, as discussed in \cite{VSPP95}, shows little spatial
variation (except near star formation sites where it is much
larger) because $0 \le \gameff < 1$ in general for the
temperature range spanned by the simulations. In particular,
$\gameff=0$ implies an isobaric 
medium. Note that $\gameff<1$ is in agreement with the elementary fact
that in the ISM denser regions are colder in that temperature range
(\cite{myers78}). In other words, the near constancy of the pressure in
this temperature range is a consequence of thermal balance (giving
$\gameff <1$) in a medium whose density is determined by the turbulent
motions. However, this near ``pressure equilibrium'' has no effect in
confining the clouds, which are in a state of constant change. As
stated in \cite{VSPP95}, the pressure is {\it slaved} to the density
field, which in turn is determined by the turbulence.

\subsection{Super- and sub-Alfv\'enic motions and magnetic field
topology}\label{super_sub}

Another point worth discussing is the sub- or super-Alfv\'enic
character of the velocity, since it has been traditionally argued that
the motions in molecular clouds are supersonic but sub-Alfv\'enic
(e.g., \cite{shu_etal87}), although more recently it has been claimed
by Padoan \&\ Nordlund (1998) that this may not be so, but rather that
motions may generally be super-Alfv\'enic. Since in our simulations
clouds form out of the general ISM, they represent a good test to see
which of these conditions develop, rather than using some
pre-determined assumption about the sub- or super-Alfv\'enic character
of the problem (e.g., \cite{mousch87}; \cite{liz_shu89}). In
fig.\ \ref{family}c the black and white arrows respectively denote
super- and sub-Alfv\'enic velocities. It is seen that {\it inside the
cloud there are both sub- and super-Alfv\'enic velocities}. Since the
velocity field does not show large magnitude fluctuations at the
transition sites, this can only be understood in terms of a change in
the Alfv\'en speed $\va$. To see this, in fig.\ \ref{family}c contours
of the magnitude $B$ of the magnetic field are shown as well. The
contours span the range from 4 to 10 $\mu$G, in intervals of 1.5
$\mu$G. Thicker lines indicate progressively larger values of $B$. The
velocity is seen to be sub-Alfv\'enic everywhere outside the cloud.
Inside the cloud, it is super-Alfv\'enic in the lower region of the
cloud. In the upper region, the velocity is sub-Alfv\'enic due to a
combination of large $B$-values and low densities. Looking down towards
the center, another super-Alfv\'enic region is seen, due mainly to the
increase in the density, which reaches $\rho=55$ cm$^{-3}$ at the peak.
However, at the peak, the velocity becomes sub-Alfv\'enic again, this
time due to a sharp decrease in the velocity magnitude itself -- sort
of a ``stagnation point'' at the density peak. These variations are
thus due again to the collisions between the magnetized gas streams,
which push the field and cause magnetic shocks and field reversals
where the fluid shocks occur, and in general produce a rather chaotic
flow. One such reversal is seen to occur on the filament extending out
of the cloud towards the lower left corner of fig.\ \ref{family}c. This
is seen on the contours as a decrease in the field's intensity towards
the central ridge of the filament. Figure \ref{vort}, which shows the
magnetic field vectors on top of the density (gray scale) and the
vorticity (contours), shows this phenomenon more clearly.

Note that this trans-Alfv\'enic character of the motions may explain why
the density peaks are ``fat'' compared to the stream-collision sites:
the information of the presence of the collision is often able to
travel upstream of the flow. Furthermore, since the collisions are
generally oblique, it is likely that magnetosonic waves propagating
perpendicular to the collision surface may overtake the perpendicular
component of the oblique fluid motion, even if the total fluid
velocity is super-Alfv\'enic. In contrast, it is well known that in
Burgers turbulence, where no pressure is available, the density
structures are as thin as allowed by viscosity exclusively (see, e.g.,
\cite{chap_sca98}).

The advection of the magnetic field described above is 
contrary to the usual assumption that the field controls the
fluid motions, forcing the gas to flow along the field
lines. Nevertheless, it is noteworthy that the field strengths
developed by the simulations in both the intercloud medium and in the
clouds are not unrealistic. The lowest values of the field occur at
the intercloud medium, and are $\sim 1 \mu$G, while the largest values
occur in the densest regions, with values $\sim 25
\mu$G. Nevertheless, the field is quite intermittent, and does not
follow a unique scaling with the density
(\cite{PVSP95}), exhibiting significant variations within the cloud of
fig.\ \ref{family}c. Moreover, even under near-equipartition conditions
between kinetic and magnetic energy, the velocity field should be
expected to exert as much of an influence on the magnetic field as
viceversa. In \S\ \ref{B_comparison} we discuss the degree
to which this field topology is consistent with observations.

\section{Comparison with observations.}\label{comp_obs}

The scenario presented here differs significantly from the standard
view in which clouds and clumps have well defined boundaries (density
discontinuities) which separate them from a 
much more tenuous, warmer medium, as in simulations of
cloud collisions (e.g., \cite{sca_pum83}; \cite{klein_etal95};
\cite{min_etal97}), of 
cloud confinement (e.g., \cite{BM92}), quasi-static contraction (e.g.,
\cite{liz_shu89}) and stability (e.g., \cite{ebbert55};
\cite{bonnor56}; \cite{mcl_pud96}), etc. On the other hand,
\cite{scalo90} has suggested that such a standard model may be an
over-idealization stemming from the low resolutions and column density
ranges of early
studies, and proposed instead a much more complex scenario in which
clouds do not have a well-defined identity. Our results support this
view, and in fact such poorly-defined identity is reflected in the
arbitrariness of cloud- and clump-finding algorithms
(e.g., \cite{will_etal94}; \cite{VSBPR97}). It is
necessary, however, to test whether the scenario proposed 
here is consistent with different observations.

\subsection{Comparison between observed spectral-line maps and
mass-weighted velocity histograms} \label{spectra}

The colliding streams we have described as responsible for cloud
formation should manifest themselves as multiple peaks in density-weighted
velocity histograms of one velocity component, the equivalent in our
simulations of the optically thin line profiles in observational
spectral line maps.  Although frequently Gaussian or other smooth
curves are used to fit such lines, conferring the idea that the lines
are uni-modal and correspond to well-defined, isolated entities, in
actuality the profiles generally exhibit multiple peaks (for recent
examples, see, e.g., \cite{falg_etal94} etc.)
which are in fact normally interpreted as ``clumps'' (e.g.,
Williams, Blitz \&\ Stark 1995, hereafter WBS95).

For comparison with observational profiles, we show in fig.\
\ref{hists} the density-weighted histogram of the $x$-component of the
velocity ($u_x$) for the ``parent'' cloud showed in
fig.\ \ref{family}b. As usual, this histogram has been obtained
using the local speeds in the cloud frame. This histogram may be
compared with the \ala{13}CO line profile for
the Rossete Molecular Cloud (RMC) shown in fig.\ 4 of
WBS95\footnote{Although fig.\ 4 in WBS95 shows CO spectra,
which are optically thick, there is little, if any, qualitative
difference with the optically thin $^{13}$CO spectra shown in fig.\ 5
of that paper}. This cloud is in many aspects comparable to the cloud of
fig.\ \ref{family}b: while the RMC's dimensions (as deduced from fig.\
17 in WBS95) are $\sim 90 \times 70$ pc, the cloud in fig.\
\ref{family}b is $\sim 250 \times 120$ pc. Furthermore, the cloud in
fig.\ \ref{family}b has a mean density of $\sim 10$ cm$^{-3}$, while the gas
sampled in the spectrum of fig.\ 4 of WBS95 has a mean density $\sim
15$ cm$^{-3}$. 

Comparing the
histograms for both of these clouds, we note several points in
common. First, both sets of data
have FWHMs of roughly 6 \kms\ when only the main features are
considered. Second, both sets exhibit high-velocity bumps, at
several \kms\ from the centroid. Finally, and probably most
importantly, the ``main'' features 
are seen to contain substructure in both sets of plots. However, while
such features have been traditionally interpreted as ``clumps'',
in our simulations they originate from extended regions
within the cloud. 

In order to explain this phenomenon, we show in fig.\ \ref{interm} a
different representation of the parent cloud (fig.\
\ref{family}b). The isocontours
denote the density field, the arrows show the velocity field
in the frame of the cloud, and the various tones of gray
represent zones with nearly the same 
$x$-component of the velocity $u_x$. The velocities range from less
than $-8$ \kms\ to more 
than $8$ \kms. Each tone of gray indicates a velocity interval of 2
\kms. Several issues are then observed. First, we note that the various
peaks in the histogram of fig.\ \ref{hists} contain contributions
from extended and separate regions in space. 
Second, local density peaks often encompass several
velocity intervals, in a similar fashion to the situation for the
entire cloud, suggesting a kind of self-similarity. Note then that the
relative height in the histogram depends mainly in the mass fraction of the
gas with the corresponding value of $u_x$,
since every velocity interval appears to sample a wide range of densities.
Third, a given feature (peak) in the histogram may have contributions
from regions with very different total
velocity vectors, which only happen to have a similar
$x$-component. 
The above results suggest that the identification of ``clumps'' as
features in velocity space is risky at best, and possibly erroneous,
as already pointed out by \cite{adler_rob92}.

A final word concerning the velocity field is in order. As pointed out
in \S\ \ref{topology}, significant vorticity is found within the
clouds. Observationally, this may be detected as velocity gradients
along the cloud's projected area on the sky, which indeed have been
observed (e.g., \cite{arq_golds86};
\cite{goodman_etal93}). Furthermore, \cite{MDJF98} have recently
reported possible direct evidence for vorticity in clouds.

\subsection{Magnetic Field: alignment, advection and field
reversals} \label{B_comparison}

Concerning the magnetic field topology, the field reversals  and
bendings mentioned in \S\ \ref{super_sub} might seem contrary to the
widely accepted fact that the magnetic field lines projected on the
plane of the sky, as obtained in polarization analysis (see, e.g.,
\cite{goodman_etal90}), are relatively smooth over the area of the
clouds. However, on closer inspection of the field topology in our
simulations, there is no contradiction. As can be seen in
fig.\ \ref{vort}, the magnetic field bendings are correlated with
bendings of the density features, so that, along elongated density
features, the field appears rather smooth.  Besides, the projection
along the line of sight can make the field lines appear smoother than
they are (\cite{BPVS98}). On the other hand, evidence that the
interaction of a shock with dense molecular gas is producing a reversal
in the line-of-sight field component in the Orion/Eridanus region has
been recently reported (\cite{heiles97}).

Note also that it is by now well established that the field lines do
not in general show a unique direction of alignment with the density
features (\cite{goodman92}). In our simulations this is also true.
Although in particular for the case shown in fig.\ \ref{vort} the field
lines are mostly parallel to the density features, in other cases they
are perpendicular (see, e.g., the concave region at coordinates $x
\sim 605$ and $y \sim 510$ in fig.\ \ref{vort}). 

\subsection{Lifetimes of turbulent clouds} \label{lifetimes}

An important consideration is whether the turbulent-fluctuation nature
of the clouds proposed  here gives cloud lifetimes consistent with
current estimates. First of all, we should point out that clouds are
not necessarily completely disrupted by the turbulence. Frequently
they are simply sheared, split, merged with others, or they can just
re-expand into the general medium if they are not very stronly
self-gravitating, again indicating that the medium is not in precise
pressure balance (see the videos mentioned above).

In any case, a conventional estimate for a typical ``lifetime''
can be given bt $\tau=l/\Delta v$, where $l$ is a characteristic size
of the cloud and $\Delta v$ its internal velocity dispersion. As an
example, we calculate this for the ``daughter'' cloud of fig.\
\ref{family}c. We find $\Delta v=2.3$ \kms, and $l\sim 30$ pixels $=
37.5$ pc. Thus, $\tau \sim 1.6 \times 10^7$ yr. This value is
consistent with various estimates of GMC lifetimes of order a few
$\times 10^7$ yr (\cite{bash_etal77}; \cite{bli_shu80};
\cite{larson81}; \cite{blitz94}). 

\section{Discussion} \label{discussion}

\subsection{Applicability to molecular gas and the role of $\gameff$}
\label{gamma}

The analyses from the previous sections have been performed on
simulations which have reasonably realistic cooling functions and a
diffuse heating which decays with increasing density (mimicking
self-shielding) (\cite{PVSP95}). As seen from eq.\ (\ref{table_gamma}),
such heating and cooling rates imply effective polytropic exponents
typically smaller than 1. These heating and cooling rates are most
appropriate for diffuse HI structures, which may be very large (up to
$\sim 1$ kpc, although small-scale structure must exist there as
well). The question then 
arises as to whether the dynamical scenario we have proposed here is
applicable to molecular clouds (MCs) and their substructure (with sizes $\sim
10$ to $\lesssim 0.1$ pc).
Although the ultimate test for this problem will require new numerical
simulations, which will be presented elsewhere, we can discuss the
issue to a certain extent here, based on recent results on the role of
$\gameff$ (\cite{tohline_etal87}; \cite{VSPP96};
\cite{PVS98}), which is a measure of the thermal characteristics of
the flow.

The main difference between the parameters of the simulations
presented here and another set 
more appropriate for MCs is the effective polytropic exponent
of the flow. (Note that the only scale-dependent quantities in our
simulations are the heating and cooling rates. All other quantities are
scale-free.)  While in general in our simulations $0 \le \gameff <
0.5$ (c.f.\ \S\ \ref{simulations}), it is possible that in molecular
gas $\gameff \gtrsim 1$ (\cite{myers78}). Actually, the situation at
densities $\gtrsim 10^3$ cm$^{-3}$ is rather uncertain, with $\gameff$
possibly being $\sim 1$ between $10^3$ and a few $\times 10^4$
cm$^{-3}$ because of the importance of collisional deexcitation
and radiative trapping in CO lines, and either larger or smaller than
unity at larger 
densities depending on the presence of embedded stellar sources and
other factors (\cite{scalo_etal98}).

In any case, we can discuss the expected effect of the gas having
$\gameff \gtrsim 1$ on the nature of the density fluctuations arising in
the flow. It has recently been shown by Passot \&\ \VS\ (1998) that at larger
$\gameff$ the density peaks are spatially more extended and of smaller
amplitude, while the density minima are very deep. At small $\gameff$
(typically $0 \le \gameff \le 1$), the situation is reversed, the peaks
being very concentrated and large, while the minima are not as deep. This is 
because the 
sound speed $c$ scales with the density as $\rho^{(\gameff-1)/2}$,
causing the peaks to re-expand faster at larger $\gameff$.
At $\gameff=1$ a critical case occurs, the fluctuations of $\ln \rho$
being symmetrical with respect to the mean. \cite{VSPP96} have calculated
the density jump for shocks in a barotropic fluid parameterized by
$\gameff$. 
Thus, if $\gameff$ is larger at MC densities than in our simulations, we
can expect the density peaks to be wider and less pronounced. But
we do not expect the general scenario of clumps being formed by
turbulent compressions to change significantly. The colliding streams,
if supersonic (or super-magnetosonic), create non-stationary
shock-bounded slabs, which simply re-expand (if not made
self-gravitating by the compression--see below) on different timescales
depending on $\gameff$ after the compression subsides. If the
compressions are subsonic no shocks form, but, for sufficiently small
$\gameff$ and/or cloud masses close to their Jeans mass, collapse may
still be induced by the compressions (\cite{tohline_etal87}).

One crucial issue, however, is whether the compressed slabs can become
self-gravitating upon the compression, and, if so, whether a
quasi-hydrostatic configuration can be formed. The former problem has been
addressed by a number of authors (\cite{hunter79};
\cite{hunt_fleck82}; \cite{tohline_etal87}; Vishniac 1983, 1994;
\cite{elm93b}; \cite{VSPP96}). Two main issues are at play
here. One is the total gravitational ($\Eg$) and internal ($\Ei$)
energies of the 
cloud (for the purpose of this discussion the ``internal'' energy can
be generalized to include all forms of energy that provide support for
the cloud against gravity). The other is the {\it rate} at which each
energy increases upon compression. If $|\Eg| > \Ei$, the cloud is
gravitationally unstable, and collapses upon a perturbation. This is
essentially an integral version of the Jeans criterion (e.g.,
\cite{bona_etal87}). However, for three-dimensional collapse, it is
well-known that, if $\gameff > \gamcr \equiv 4/3$, then the collapse
will ulimately be halted, since $\Ei$ increases faster
than $|\Eg|$. The corresponding values of $\gamcr$ for
2- and 1-dimensional collapse are respectively 1 and 0
(\cite{VSPP96}). 

Conversely, if a cloud is initially stable according to Jeans ($\Ei
\ge |\Eg|$), it may
be made unstable upon compression. Tohline et al.\ (1987) have
calculated the required Mach number of a 3-dimensional compression as
a function of $\gameff$ for this to occur, provided that $\gameff <
4/3$ (for $0 < \gameff <1$ they found that the required Mach number is
independent of the cloud's mass). Two- and one-dimensional
compressions require $\gameff < 1$ and 
$\gameff < 0$, respectively (\cite{VSPP96}). The issue is then the
following. In order for an initially stable cloud to be pushed over
the internal energy ``barrier'' (\cite{tohline_etal87}) by a strong
enough turbulent
compression, it is necessary that $\gameff < \gamt$, where
$\gamt$ is the critical $\gamma$ for $|\Eg|$ to increase faster
than $\Ei$ upon the turbulent compression. From there on,
gravitational collapse takes over. Since in general turbulent
compressions are expected to be less than three-dimensional, then
generally we expect $\gamt \le \gamg$, where
$\gamg$ is the critical $\gamma$ for gravitational
collapse.\footnote{Here we are allowing for the possibility of
gravitational collapse to occur faster in one direction than in
the other two, as is the case of cosmological ``pancakes''.} Then,
$\gameff < \gamt \le \gamg$. Thus, we conclude that,
if collapse was triggered by a turbulent compression, it cannot later
be halted by the internal energy, unless $\gameff$ changes in the
process. The latter may happen in the late stages of collapse, if
the dynamical timescales become shorter than the radiative cooling
timescales, causing $\gameff$ to approach the actual heat capacity
ratio of the gas $\gamma$, rather than the value derived from the
condition of thermal equilibrium (see, e.g., \cite{PVSP95}).

We should insist here that the effects of all
forms of support, such as the magnetic pressure and turbulent support,
can in principle be included in $\gameff$. On the other hand, this is
a very simplistic description, since in general $\gameff$ may be a
function of the density, and the anisotropic nature of magnetic
support may not be well described by a simple polytropic exponent. In
any case, this discussion illustrates our main point: {\it
quasi-hydrostatic configurations cannot be produced out of turbulent
fluctuations, unless the flow becomes closer to adiabatic
than to isothermal during the collapse}. Fluctuations either rebound
or collapse. However, 
at small $\gameff$, the rebound time is long (\cite{VSPP96}; \cite{PVS98}).

We can examine more quantitatively the conditions under which adiabatic
compression or collapse would occur if thermal pressure provides the main
support mechanism for a clump,
by comparing the cooling timescale $\tau_{\rm cool}=nkT/\Lambda$ ($\Lambda$ =
cooling rate in units of erg
cm$^{-3}s^{-1}$) to the compression time $\tau_{\rm comp}\equiv L/v$ and the
collapse time
$\tau_{\rm coll}\equiv(G\rho)^{-1/2}$.  For the cooling rate we distinguish two
density regimes.  For
$10^3\lesssim n \lesssim 10^4$ cm$^{-3}$ the cooling is dominated by CO
line cooling.  We 
adopt a rough approximation for
the CO cooling rate (see references in Scalo et al.\ 1998) as
$\Lambda\sim10^{-26}n_3T^2$, where
$n_3=n/10^3$cm$^{-3}$ and we have assumed that collisional deexcitation
yields a linear (rather than
quadratic) density dependence at these densities.  We recognize that this
is a crude representation of the
molecular cooling rate, but it suffices for our order of magnitude
estimates.  At larger densities the
cooling will be dominated by gas-grain collisions (if there are no embedded
protostellar sources), with a
rate $2\times10^{-33}n^2T^{1/2}\ (T-T_{\rm gr}$), where $T_{\rm gr}$
is the grain temperature and the coefficient
depends somewhat on the adopted grain parameters.  We assume
$T-T_{\rm gr}\approx T$ for our order of magnitude
estimates.
 
For compressions, we find that the condition for adiabatic evolution
($\tau_{\rm comp}<\tau_{\rm cool}$) yields a
critical length scale $L_{\rm pc}=L/1$pc
\begin{displaymath}
L_{\rm pc}< \left\{4\times10^{-3}M/T_{10}^{1/2}\ \ \hbox{CO cooling}\atop
7\times10^{-4}M/n_5\ \ \
\hbox{gas-grain cooling,}\right.
\end{displaymath}
where $M$ is the Mach number of the compression, $T_{10}=T/10$K, and
$n_5=n/10^5$cm$^{-3}$.  For any
reasonable Mach number adiabatic compression is only possible for tiny
regions of size $\lesssim 10^{-2}$ pc.
 
If a (non-adiabatic) compression results in collapse, the condition for
adiabatic collapse is
$\tau_{\rm coll}<\tau_{\rm cool}$.  At densities $10^3\lesssim
n \lesssim 10^4$, the adopted CO cooling rate results in the
condition $n_3 > 2.1 \times10^3 T_{10}^2$.  But at such large densities the
cooling will be controlled by
gas-grain collisions, not molecular line cooling.  Using the gas-grain
cooling rate, the condition for
adiabatic collapse is found to be $n_5 < 1.2 \times 10^{-3}T_{10}^{-1}$. But at
such small densities molecular
line cooling would dominate and be non-adiabatic.  Thus for collapse
induced by compression, adiabatic
evolution will never occur until the collapsing entity is so dense that it
becomes opaque to its own
infrared radiation, which only occurs at very large (protostellar) densities.
 
These estimates indicate that turbulence-induced compressions or collapse
will generally be controlled by
the effective polytropic index $\gameff$ set by the thermal
equilibrium condition, since $\tau_{\rm cool}$
is smaller than the relevant dynamical timescales, except for extremely
small scales of compression or at
protostellar densities for collapse.  We conclude that quasi-static
configurations are unlikely over a very
wide range of physical parameters.

Is there any observational
evidence for a dynamical rather than quasistatic state of
clumps? A preliminary answer appears affirmative. Indeed,
\cite{girart_etal97} have recently reported on a system which 
may be interpreted as a collision of streams in L723. Furthermore,
\cite{tafalla_etal98} have reported a core with inward motions that do
not correspond to classical gravitational infall models. A detailed
comparison of these data with our scenario will be presented elsewhere.

\subsection{Effect of dimensionality}

In this paper we have used two-dimensional simulations, with the known
risks involved by them, such as a different  distance dependence of
gravity, the absence of a magnetic dynamo, and the possibility of
reverse cascades (although this is probably less important in the
compressible case, where the energy spectrum may be determined by the
shocks rather than by cascade processes -- see, e.g., \cite{VS98}). This is a
necessary price to pay for the larger resolution allowed by the
two-dimensionality, which is essential for the type of hierarchical
structure analyses performed here.

However, we do not expect the
two-dimensionality of our simulations to be a matter of concern for the
validity of our results. The main points raised here are those of the
mechanism of cloud formation and its consequences regarding the mass
flux through cloud boundaries, the deformation of the magnetic field by
the flow, and the interpretation of line profile observational data.
These do not seem to depend on the dimensionality of the problem,
since they originate from the shocks and collisions of streams, which
occur in both two and three dimensions. In any case,
intermediate-resolution 3D simulations are currently being performed
which will allow us to test these expectations.

\section{Summary and conclusions}\label{conclusions}

In this paper we have conducted a close examination of the topology of
the density, velocity and magnetic fields that result in our 2D
simulations of the ISM at the kpc scale, in particular within the
clouds that emerge as the turbulent density fluctuations in that
medium. We found that, in general, the cloud boundaries are rather
arbitrary, being defined only by a density threshold criterion, but not
by a physical boundary, such as a density or velocity discontinuity. The same
arbitrariness is likely to be present in clouds defined by the tracer
used to observe them, since a cloud may seem to ``end'' where the
tracer density falls below a threshold density necessary to excite the
transition, but this need not correspond to any abrupt drop in the mass
density.

We furthermore considered a hierarchy of clouds in the simulations, by
using a sequence of three progressively larger density thresholds for
defining them. For clouds at all levels of this hierarchy, the velocity
field in the simulations is continuous across the clouds'
``boundaries'', indicating that the clouds are being formed by
colliding gas streams (\cite{hunter_etal86}; \cite{elm93b};
\cite{VSPP95}), with the density increasing towards the collision site.
The density enhancements need not be confined exclusively to the
collision site as in Burgers flows, because the velocities involved are
typically trans-Alfv\'enic, meaning that the information on the
collision can often propagate upstream (see also \cite{pouq_etal91}).
Furthermore, since the collisions are generally oblique, it is likely
that magnetosonic waves propagating perpendicular to the collision
surface may overtake the oblique fluid motion, even if the total fluid
velocity is super-Alfv\'enic.  This scenario is in sharp contrast with
static models in which the clouds are immersed in an intercloud medium
of lower density but higher temperature, and confined by pressure
balance between the two ``phases'' (e.g., \cite{BM92}; \cite{MZ92};
\cite{mcl_pud96}). In our simulations both ``phases'' exist, but the
transition between them is smooth, and the situation, rather than
 being static, is highly dynamic.

To further characterize the role of the flux across the clouds'
boundaries, we considered the surface ($\xruu$) and volume ($\ekin$)
kinetic terms in the virial theorem for the whole ensemble of clouds at
one temporal snapshot in the simulation. It was found that both terms
are typically of the same order of magnitude, indicating that the
surface terms contribute an amount comparable to the total kinetic
energy of the clouds to their overall virial balance. This was
interpreted as a signature of the fact that, when the clouds are formed
by stream collisions, the same flow is responsible for both the total
kinetic energy contained within the clouds and for the flux across its
boundaries.

We discussed the differences between the thermal and turbulent
pressures, emphasizing the fact that while the former is microscopic
and isotropic, the latter may involve macroscopic scales, comparable or
even larger than the clouds' sizes, and is in general anisotropic, as
indicated by its off-diagonal contribution to the total pressure tensor
$\Pi_{ij}$. We thus concluded that turbulent pressure confinement
is a contradictory concept, because it
involves motions at large scales which by definition cause a change in
the clouds' boundaries. In turn, we argued that thermal pressure
equilibrium in a non-static medium is irrelevant, because inertial
motions may still distort or disrupt a cloud, even under thermal
pressure balance with its surroundings.

Additionally, we studied the magnetic field topology and strength in
the clouds in the simulations, and their implication on whether the
motions are super- or sub-Alfv\'enic, finding both super- and
sub-Alfv\'enic motions within the clouds, supporting the suggestion
that the field is highly intermittent (\cite{PN98}). We noticed that
the field is significantly distorted and advected by the flow, in
agreement with the result that the flow within the clouds is
trans-Alfv\'enic. The field tends to be correlated with the density
features, although without a unique mode of alignment (\cite{PVSP95}).
This causes the field to look relatively smooth along density features,
in agreement with most polarization studies (\cite{goodman_etal90}),
but at the same time being bent where the density filaments are bent as well. 

A crucial question is whether the scenario depicted in this paper is
consistent with observational data. We addressed this problem at the
level of density-weighted
velocity histograms, which can be compared to observational spectral
line profiles from real clouds of similar characteristics to those in
the simulations. Although there are limitations to this comparison, our
velocity histograms are encouragingly similar to the observational line
profiles, both exhibiting comparable FWHMs ($\sim 6$ \kms) and
high-velocity isolated features, and both containing not completely
resolved substructure, indicating multiple components. We emphasized
that while this substructure is usually interpreted as isolated
``clumps'', in our simulations it originates from different
regions of the cloud, which may correspond to different ``streams''
and in general have one similar component of the velocity but very
different total velocity vector, as found also by \cite{adler_rob92}.
We also mentioned observational evidence for field reversals in shocks
(\cite{heiles97}), and noted that the topology of the magnetic field
seen in the simulations is consistent with observations
(\cite{goodman92}).

Finally, we discussed at considerable length the possibility that the
scenario of clouds as turbulent density fluctuations may apply to
molecular clouds and their substructure (clumps and
cores). Essentially, only the adopted cooling and heating functions in
our simulations distinguish them from a case more appropriate to molecular
clouds, which are believed to behave in a nearly isothermal
manner. Based on recent results on the effect of the effective thermal
behavior (\cite{tohline_etal87}; \cite{VSPP96}; \cite{PVS98}),
parameterized by the effective polytropic exponent $\gameff$, we
conjectured that only the shape (height and width) of the density
fluctuations is expected to change, but not the general character of
the inflowing streams characteristic of their turbulent nature. We
also suggested that quasi-hydrostatic configurations cannot be
produced from turbulent compressions, and that density fluctuations
must either collapse or rebound, unless the effective thermal behavior
becomes closer to adiabatic during the collapse ($\gameff \ge
4/3$). However, for small
$\gameff$, the rebound times can be quite long. We demonstrated, using
appropriate cooling rates, that adiabatic evolution, and hence
quasistatic configurations, are unlikely to occur either during
compressions or during compression-induced collapse.

In summary, we conclude this paper by stating that the results
presented here are not in contradiction, but rather quite
consistent, with observational data. They seem to be, however, discrepant
with the standard interpretation of those data in terms of sharply-bounded
clouds in pressure equilibrium with their surroundings, and with the
identification of velocity features in spectral-line data with
isolated ``clumps''. It is interesting that our results give theoretical
support to the suggestion made long ago by \cite{chandra_munch52}
that the interstellar medium may not best be visualized as a system of
{\it discrete} clouds.  As they pointed out, the ``cloud model'' is
difficult to 
displace because, as they say, ``...a tendency to argue in circles can be
noted in the literature, in that confirmation for the {\it picture} of
interstellar matter as occurring in the form of discrete clouds is sought
in the data analyzed''.

\acknowledgements

We thank Alyssa Goodman, Thierry Passot, Luis Rodr\'iguez
and Jonathan Williams for carefully reading early versions of the
manuscript and 
providing us with detailed comments. We also thank Paola D'Alessio and
Susana Lizano for useful discussions. We acknowledge the hospitality of
the star formation group at the Harvard-Smithsonian CfA. The
simulations were performed on the Cray Y-MP/4-64 of DGSCA, UNAM. This
work has received partial support from grants UNAM/DGAPA IN105295 and
UNAM/CRAY SC-008397 to E.V.-S.\ and a UNAM/DGAPA fellowship to J.B.-P.

\begin{figure}
\plotone{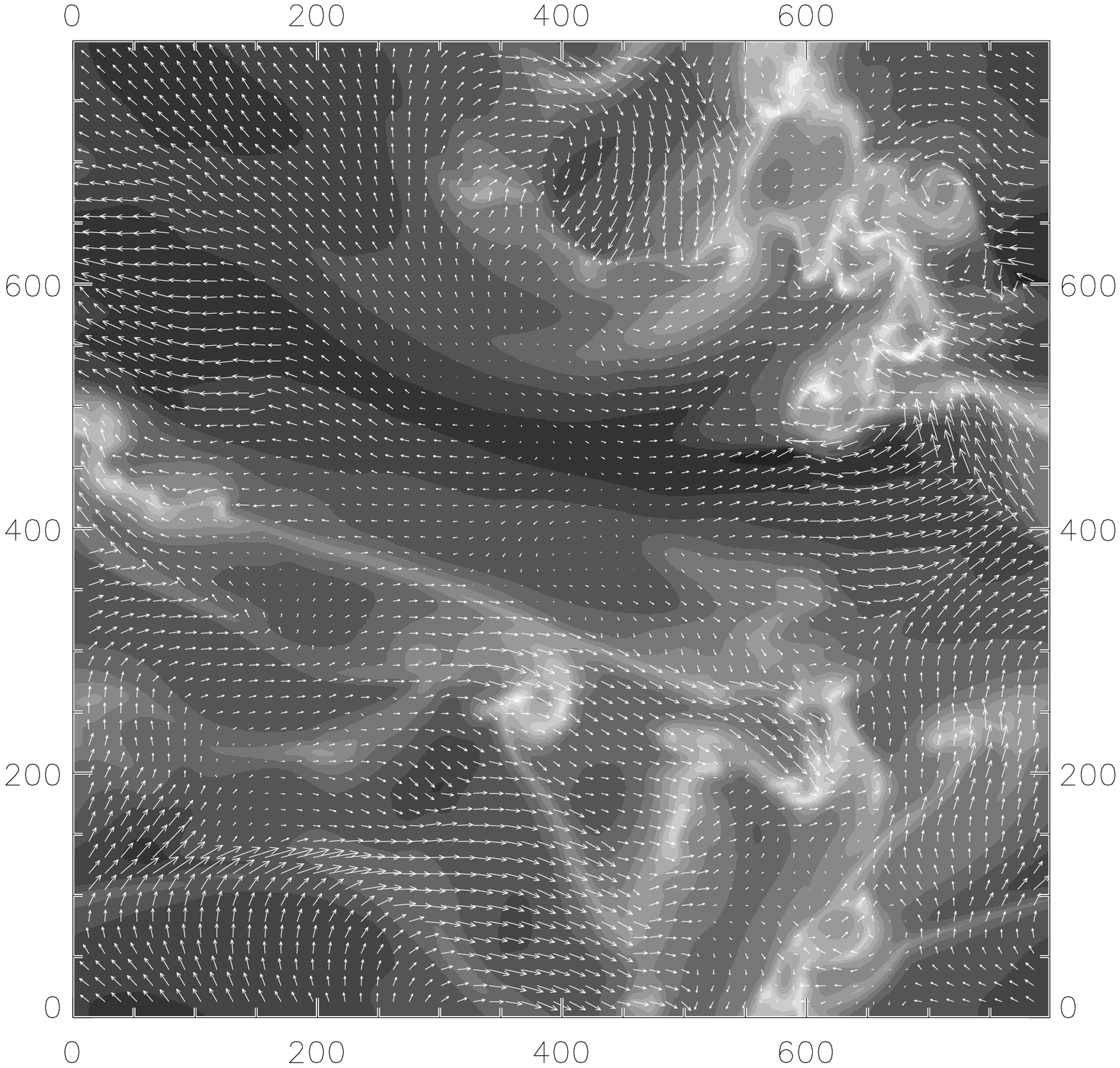}
\end{figure}

\begin{figure}
\plotone{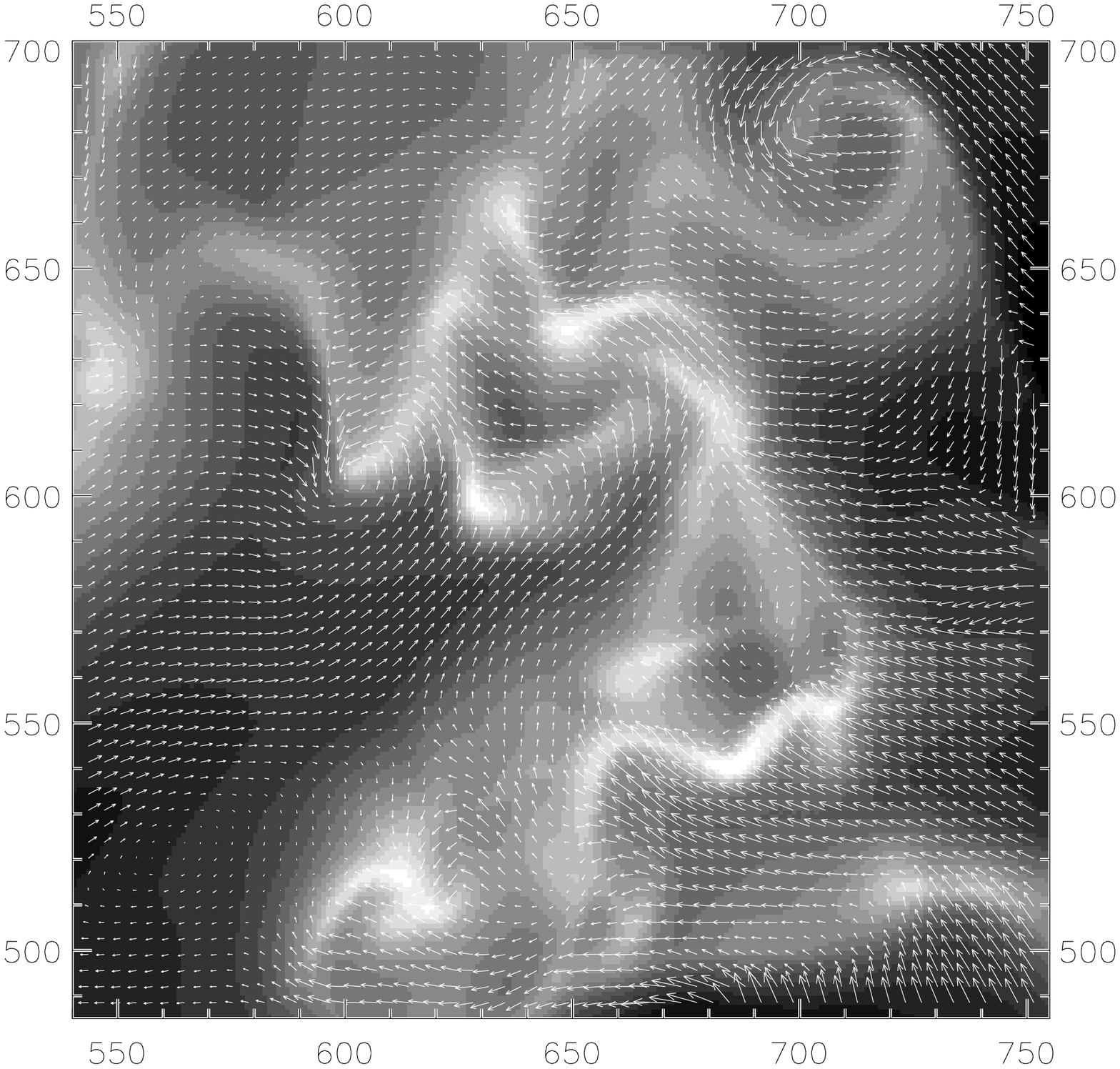}
\end{figure}

\begin{figure}
\plotone{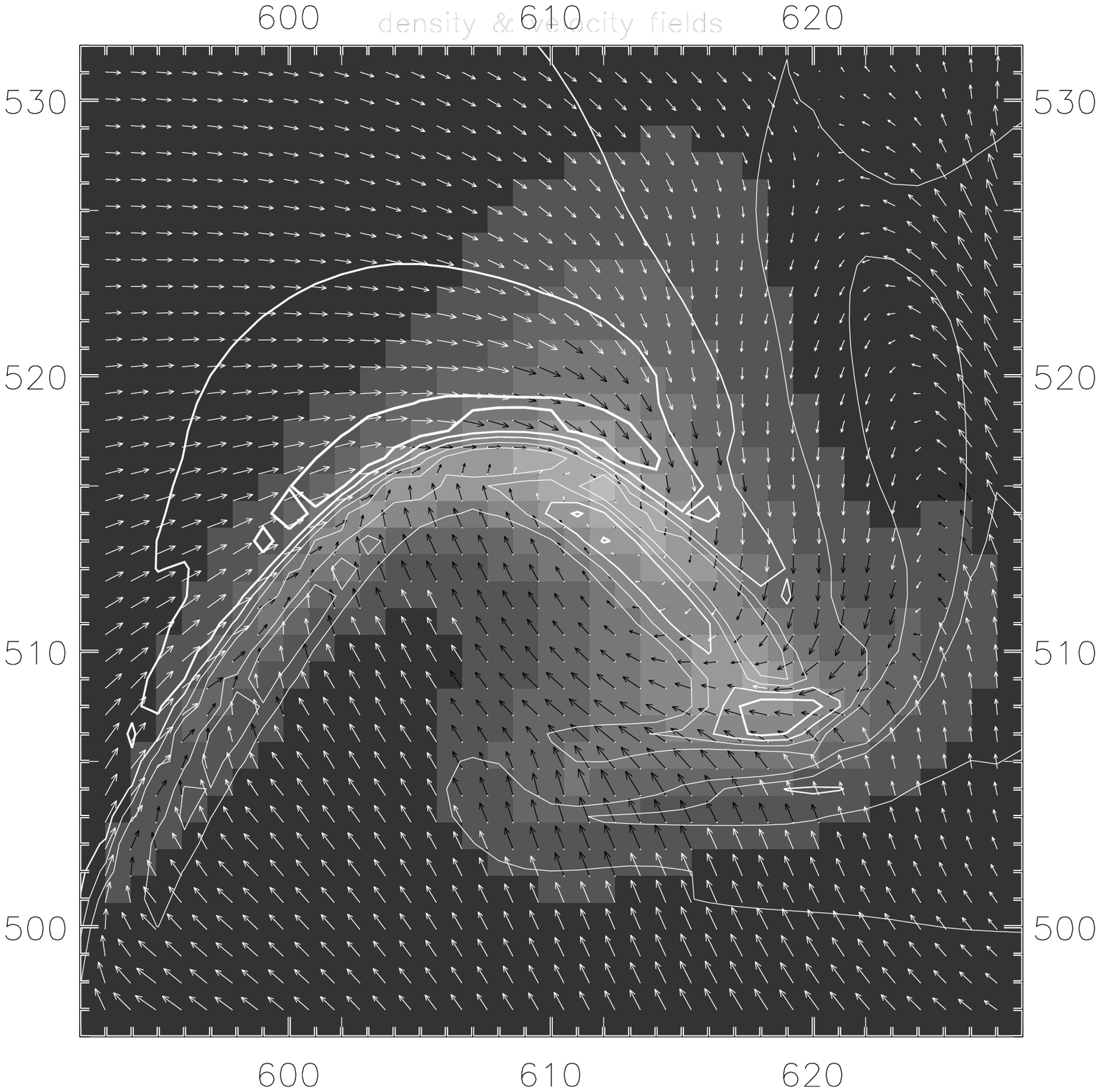}
\caption{Three hierarchical levels of clouds in the simulation,
showing the logarithmic density scale with shades of gray, and the
velocity field with arrows. Pixel numbers are indicated by the
grids. a) The whole simulation field field, 800 
pixels on a side, corresponding to 1 kpc. The gray scale shows structures
with $\rho \geq \rhot=1.5$ cm$^{-3}$ b)
Magnification of the complex that appears when setting $\rhot=4$ in
the upper right quadrant of fig.\ 1a. c) 
Magnification of the cloud that appears when setting $\rhot=8$ in the
lower left region of fig.\ 1b. Black arrows indicate super-Alfv\'enic 
velocities in the frame moving with the cloud, and white arrows
indicate sub-Alfv\'enic velocities. The contours give the magnetic
field strength $B$. Thicker countours indicate larger $B$-values. The
maximum density value is 55 cm$^{-3}$.
\label{family}}
\end{figure}

\begin{figure}
\plotone{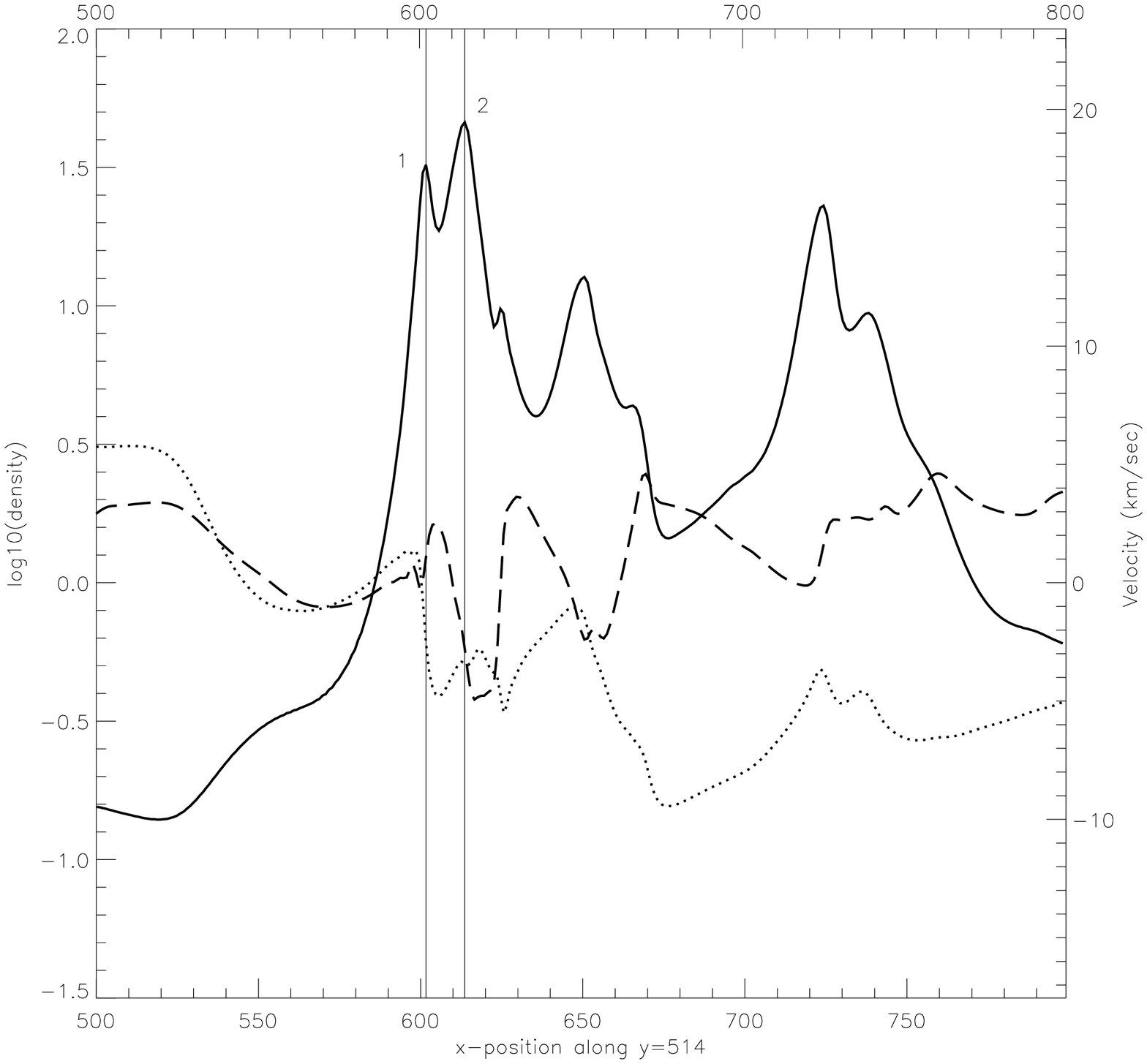}
\end{figure}

\begin{figure}
\plotone{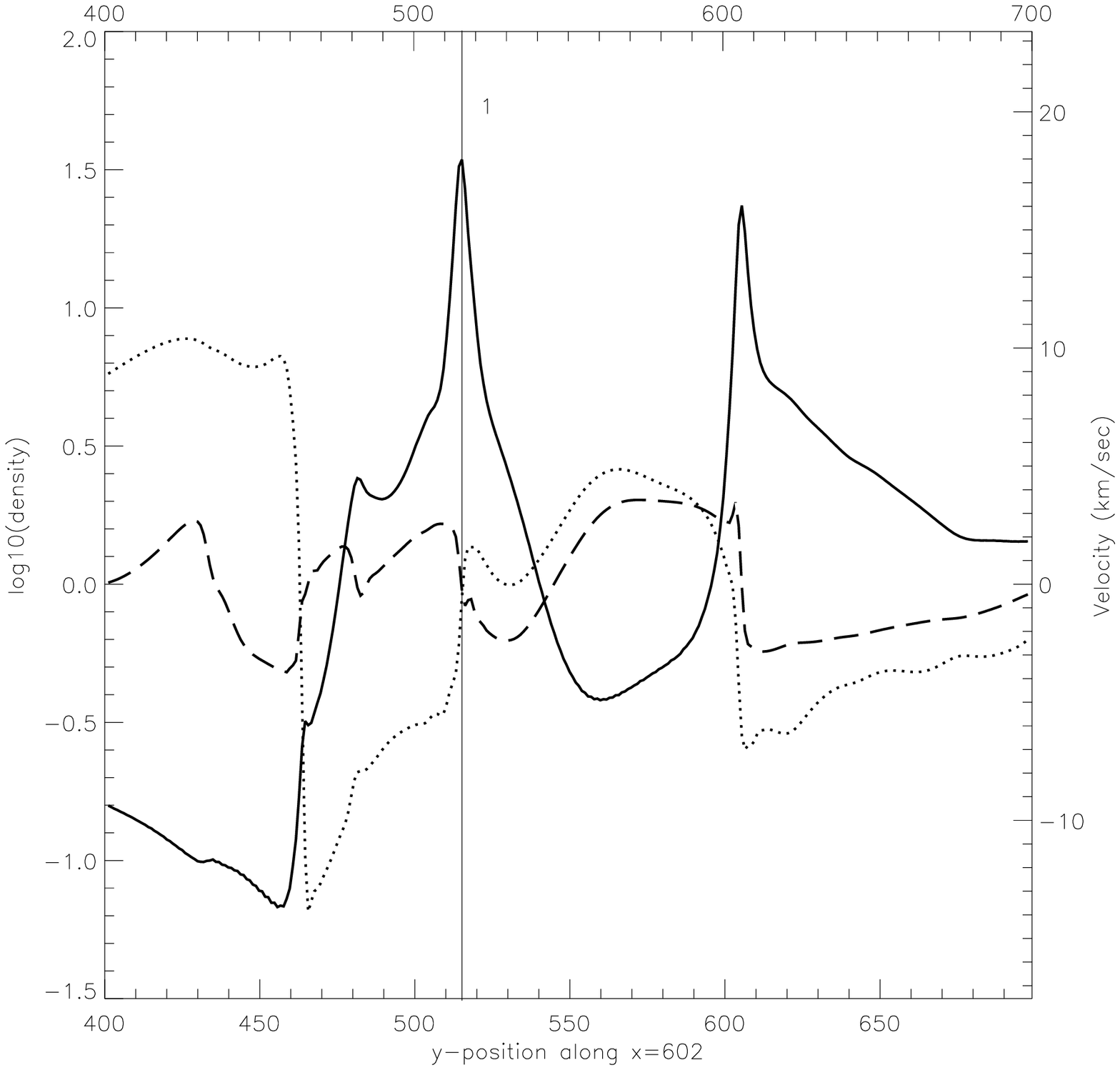}
\end{figure}

\begin{figure}
\plotone{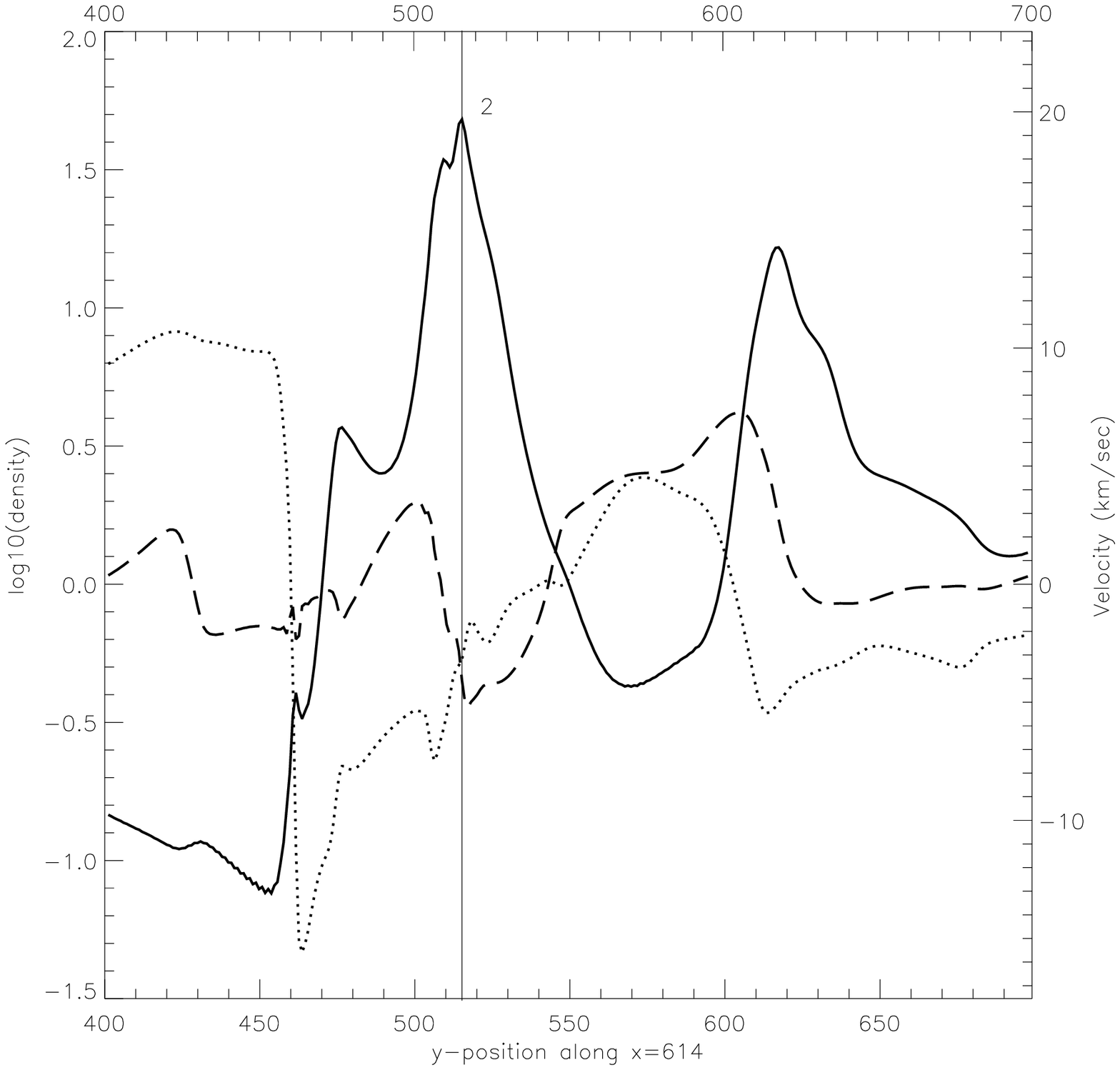}
\caption{Cuts passing through the cloud shown in fig.\ \ref{family}c,
(as indicated by the lines in fig.\ \ref{div}), showing the log of the
density (solid line), and the $x$- (dotted line) and $y$- (dashed line)
components of the velocity. a) Cut along $x$ at $y=514$. b) Cut along
$y$ at $x=602$. c) Cut along $y$ at $x=614$. In a), two peaks are
indicated. Peak 1 is seen to correspond to an $x$-shock (abrupt
negative gradient) in a) and to a
$y$-shock in b). Peak 2 is almost imperceptible in $u_x$, but is seen
to correspond to a $y$-shock in c). Also, note the ``plateau'' in $u_x$
extending 50 pixels to the right of peak 1, with average negative
values, although with further substructure. Note that the scale is
larger than in fig.\ \ref{family}c, in order to give a broader view of
the velocity components.
\label{cut}}
\end{figure}

\begin{figure}
\plotone{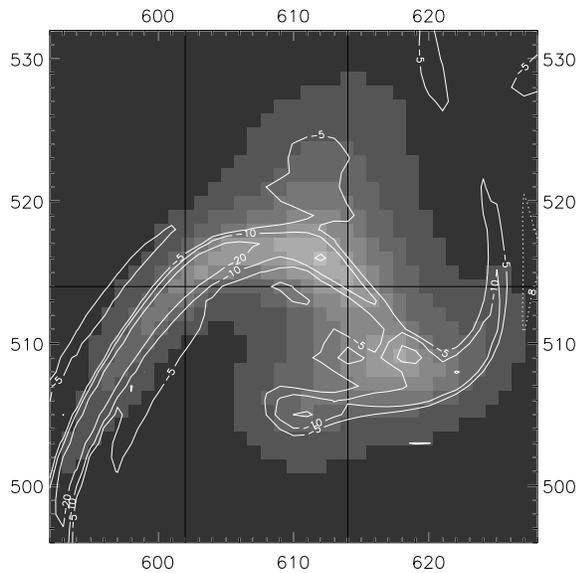}
\caption{The same cloud as in fig.\ \ref{family}c, but with superimposed
velocity divergence contours. Density maxima are
seen to correspond to maximum negative values of the divergence. Also
indicated are the lines along which the cuts of fig.\ \ref{cut} are taken.
\label{div}}
\end{figure}

\begin{figure}
\plotone{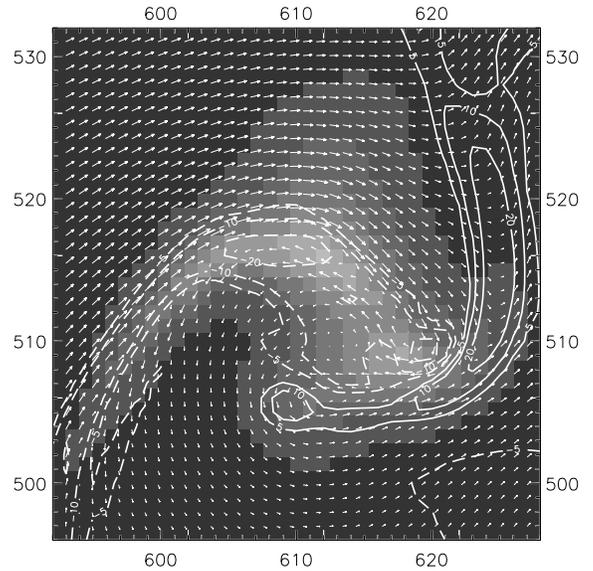}
\caption{Same cloud as in fig.\ \ref{family}c, but with superimposed
vorticity contours. There is a significant correlation between the
density (and therefore the divergence) and the vorticity fields. We
interpret this as a result of the generally oblique nature of the
stream collisions. Solid (dashed) contours indicate positive
(negative) vorticity. Also shown are the magnetic field vectors (arrows). Note
the correlation between density and magnetic field bendings and
the field reversal along the central ridge of the cloud. Note also the
general alignment of the field with the density, although at many
places the field crosses the cloud's ``boundary'' perpendicularly (e.g.,
at the concave region below the left filament).
\label{vort}}
\end{figure}

\begin{figure}
\plotone{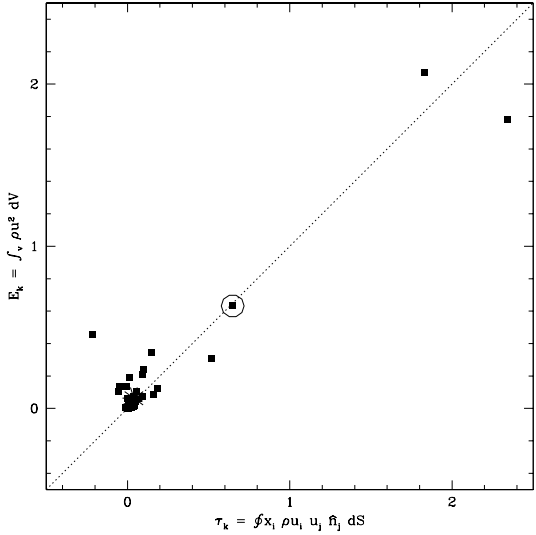}
\caption{Kinetic energy $\ekin$ vs. the kinetic surface term $\xruu$
for all clouds with areas larger than 300 pixels in the whole
field. The complex of fig.\ \ref{family}b is shown by a circle, and
the cloud in fig.\ \ref{family}c is shown by the star. Note the
similarity of the two terms for most of the clouds,
indicating that both contribute in similar amounts to the clouds'
virial balance.
\label{kinetic_terms}}
\end{figure}

\begin{figure}
\plotone{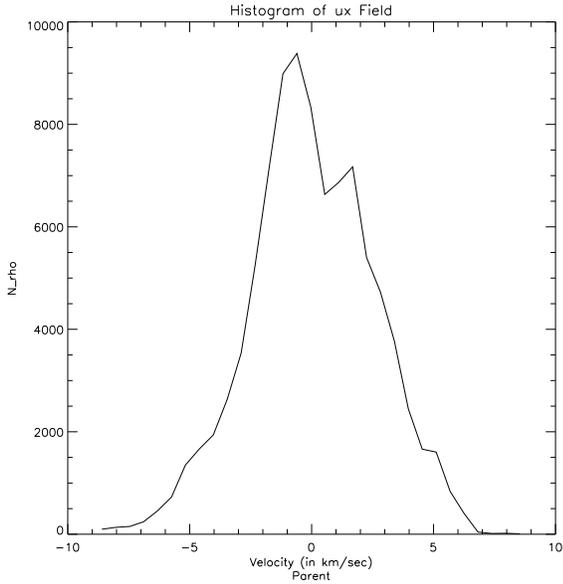}
\caption{Density-weighted velocity histograms for the parent cloud (fig.
\ref{family}b). Note the FWHM $\sim$5--7 km s$^{-1}$, and the
multi-component structure. Also, note the low-decaying wings.
\label{hists}}
\end{figure}

\begin{figure}
\plotone{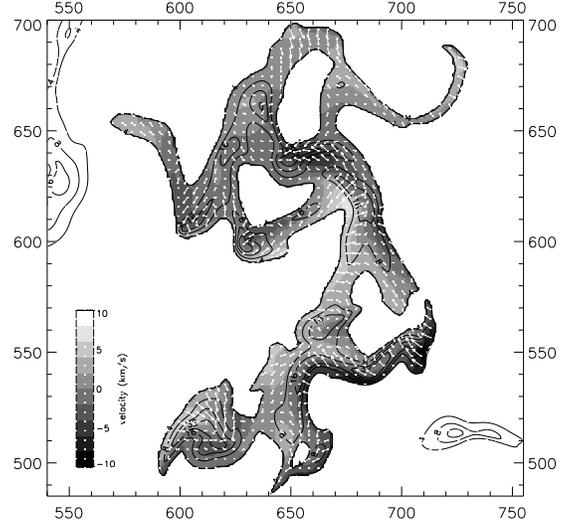}
\caption{Same cloud as in fig.\ \ref{family}b, but indicating 
the magnitude of the local $x$-component of the velocity (gray-scale),
in order to identify the origin of the $x$-velocity features appearing in the
histogram of fig.\ \ref{hists}. The velocity ranges are indicated in
the sidebar. The arrows and contours indicate the velocity and the
density fields, respectively. Three points are noticed. 1. The
contribution to any given velocity interval originates from extended
regions throughout the cloud. 2. Regions contributing to a given
velocity interval contain zones of significantly different
densities. 3. Regions with the same 
$x$-component of the velocity can have completely different
total velocity vectors.
\label{interm}}
\end{figure}

\end{document}